\documentclass[aps,prl,twocolumn,tightenlines,floatfix,nofootinbib,amsmath,amssymb]{revtex4-1}

\usepackage{bm}
\usepackage{amsfonts}
\usepackage{latexsym}
\usepackage[latin1]{inputenc}
\usepackage{graphicx}
\usepackage{amsmath}

\usepackage{palatino}
\usepackage{mathpazo}
\newcommand{\be}{\begin{equation}}
\newcommand{\ee}{\end{equation}}
\newcommand{\bear}{\begin{eqnarray}}
\newcommand{\eear}{\end{eqnarray}}

\newcommand{\rc}{{\rm c}}
\newcommand{\rK}{{\rm K}}

\newcommand{\rgw}{{\rm gw}}

\newcommand{\cR}{{\cal R}}

\def\jnl@style{\it}
\def\aaref@jnl#1{{\jnl@style#1}}

\def\aaref@jnl#1{{\jnl@style#1}}

\def\aj{\aaref@jnl{AJ}}                   
\def\apj{\aaref@jnl{ApJ}}                 
\def\apjl{\aaref@jnl{ApJ}}                
\def\apjs{\aaref@jnl{ApJS}}               
\def\apss{\aaref@jnl{Ap\&SS}}             
\def\aap{\aaref@jnl{A\&A}}                
\def\aapr{\aaref@jnl{A\&A~Rev.}}          
\def\aaps{\aaref@jnl{A\&AS}}              
\def\mnras{\aaref@jnl{MNRAS}}             
\def\prd{\aaref@jnl{Phys.~Rev.~D}}        
\def\prl{\aaref@jnl{Phys.~Rev.~Lett.}}    
\def\qjras{\aaref@jnl{QJRAS}}             
\def\skytel{\aaref@jnl{S\&T}}             
\def\ssr{\aaref@jnl{Space~Sci.~Rev.}}     
\def\zap{\aaref@jnl{ZAp}}                 
\def\nat{\aaref@jnl{Nature}}              
\def\aplett{\aaref@jnl{Astrophys.~Lett.}} 
\def\apspr{\aaref@jnl{Astrophys.~Space~Phys.~Res.}} 
\def\physrep{\aaref@jnl{Phys.~Rep.}}      
\def\physscr{\aaref@jnl{Phys.~Scr}}       
\def\commat{\aaref@jnl{Comm.~Math.~Phys.}}		
\def\science{\aaref@jnl{Science}}		
\def\cqg{\aaref@jnl{Classical Quant.~Grav.}}		
\def\jpcs{\aaref@jnl{JPCS}}					
\def\ijmpd{\aaref@jnl{Int.~J.~Mod.~Phys.~D}}			
\def\grg{\aaref@jnl{Gen.~Relat.~Gravit.}}		
\def\rpp{\aaref@jnl{Rep.~Prog.~Phys.}}		
\def\rmp{\aaref@jnl{Rev.~Mod.~Phys.}}		
\def\jpg{\aaref@jnl{J.~Phys.~G~Nucl.~Partic.}}		
\def\npa{\aaref@jnl{Nucl.~Phys.~A}}		

\begin{document}

\title{The $f$-mode instability in relativistic neutron stars}

\author{E. Gaertig, K. Glampedakis, K. D. Kokkotas, and B. Zink}

\affiliation{Theoretical Astrophysics, IAAT, Eberhard-Karls University of T\"ubingen, T\"ubingen 72076, Germany}

\begin{abstract}

We present the first calculation of the basic properties of the $f$-mode instability
in rapidly rotating relativistic neutron stars, adopting the 
Cowling approximation.
By accounting for dissipation in neutron star matter, i.e. shear/bulk viscosity and superfluid mutual friction,
we calculate the associated
instability window.
For our specific stellar model, a relativistic polytrope, we obtain a 
minimum gravitational growth timescale (for the dominant $\ell=m=4$ mode) of the order of $10^3-10^4\,\mbox{s}$ 
near the Kepler frequency $\Omega_\rK$ while the instability is active above $\sim 0.92\,\Omega_\rK$ and for temperatures $\sim (10^9 - 2 \times 10^{10})\,\mbox{K}$, characteristic of newborn neutron stars. 

\end{abstract}

\pacs{}

\maketitle


{\em Introduction---} A classic result in gravitational physics is that pulsation modes in rotating compact stars 
can be driven unstable under the emission of gravitational radiation. This is accomplished by the so-called 
Chandrasekhar-Friedman-Schutz (CFS) mechanism \cite{chandra,cfs} which is based on the general notion of oscillations 
changing from  counter- to co-rotating (with respect to the stellar rotation) as a result of rotational dragging.

It is envisaged that CFS-unstable modes could play a key role in
neutron star astrophysics by producing detectable amounts of gravitational radiation which can be used to probe the neutron star interior structure \cite{seismo, andersson2003}. Additionally, they could provide a 
mechanism for limiting the stellar spin frequency \cite{nakk01}. The realization of such gravitational-wave `asteroseismology' 
is expected to take place with the planned Einstein Telescope or more optimistically even with the Advanced LIGO and VIRGO instruments in the near future.

The initial work on the subject of gravitational radiation-driven instabilities was focused on the 
instability of the $f$-mode in Newtonian neutron star models \cite{friedman83,il91,lm95,lai95}
and concluded that such systems might be observable sources of gravitational waves, even beyond our Galaxy,
in the form of newborn neutron stars spinning near the mass-shedding limit. 
Subsequent work in the field was mostly driven by the discovery of a similar instability in the inertial $r$-modes, and its
potential relevance for systems like the accreting neutron stars in LMXBs  \cite{nakk01}.

In more recent years, progress in numerical hydrodynamics  has allowed the modelling of rapidly rotating 
{\em relativistic} neutron stars. Several calculations \cite{ye97,ye99,sf98,morsink99,zink10} have shown that relativistic 
stars enter the $f$-mode instability region at a lower spin rate as compared to their Newtonian counterparts. 
This is an indication of an {\em enhanced} $f$-mode instability in relativistic stars, and provides strong motivation for 
a renewed investigation of the subject.  
 
With the purpose of corroborating the case of a stronger relativistic $f$-mode instability, in this Letter we present the first
calculation of the growth rate of this instability in relativistic stars and the associated `instability window', 
that is, the stellar parameter space where the mode growth due to gravitational radiation overcomes the 
dissipative action of neutron star matter.


{\em The f-mode in rapidly rotating relativistic stars---} 
We model a neutron star as a self-gravitating relativistic fluid
obeying a polytropic equation of state. The pulsations of this system are studied by
a linear time-domain numerical code which evolves the 
general relativistic hydrodynamical equations $\nabla_\mu T^{\mu\nu}=0$ in the fixed spacetime of 
the unperturbed `background' star.
A key property of our model is its ability to treat {\em rapidly} rotating stars, with the stellar angular frequency $\Omega$ only restricted by the Kepler 
mass-shedding limit $\Omega_\rK$. This numerical scheme has been recently applied to the general 
study of the oscillation spectrum of rapidly rotating (both uniformly and differentially) relativistic stars 
\cite{egkk08,egkk09,kruger10}. 

More relevant to the scope of this Letter is our recent work \cite{egkk11} which has provided state-of-the-art results on 
the $f$-mode oscillation frequency and gravitational radiation damping/growth rate as a function of rotation for a collection 
of polytropic models. Referring the reader to that paper for details on the numerical techniques, here we simply 
outline the general strategy followed. A consistent calculation of the mode's gravitational radiation-induced 
damping/growth rate would require to move beyond the Cowling approximation and consider the evolving spacetime 
itself. Since this requires very large meshes and long evolution times, and is therefore impractical at present, we content ourselves with an 
approximate solution to the problem, using the standard multipole expansion for the gravitational wave flux.
This is (e.g. \cite{il91}),
\be
\dot{E}_\rgw = \sigma_r \sum_{\ell \geq 2}  \sigma_i^{2\ell+1} N_\ell 
\left ( | \delta D_{\ell m} |^2 + | \delta J_{\ell m} |^2\right )
\label{gwflux}
\ee
where $\sigma_r$ and $\sigma_i = \sigma_r -m\Omega$ is the mode frequency measured in the rotating and inertial 
frame respectively, $\delta D_{\ell m}$ and $ \delta J_{\ell m}$ are the mode's mass and current multipole moments,
and $N_\ell$ is the usual coupling constant. 

In calculating the gravitational wave flux we use the relativistic mode eigenfunctions in the volume integral
for $\delta D_{\ell m}$ and neglect the current multipoles here. We follow a similar procedure
for calculating the mode's total energy $E_{\rm m}$ (measured in the
rotating frame). Since we apply the Newtonian expressions here, though weighted using the proper spatial volume element
on the spacelike hypersurface, we introduce a slight inconsistency into the calculation. The induced error 
should be comparable to using the quadrupole formula for gravitational wave extraction \cite{shibata},
and is expected to be small compared to other modeling uncertainties.

The timescale associated with gravitational wave damping/growth is given by 
(see, for example, \cite{il91}) $t_\rgw = 2E_{\rm m}/\dot{E}_\rgw$.
The instability is indicated by a negative $t_\rgw < 0$. In turn, this requires a $\sigma_i < 0$ 
which is equivalent to the statement that $\Omega$ should exceed in magnitude the mode's ``pattern speed'' 
$-\sigma_r/m$. This is a short description of the  CFS instability mechanism \cite{cfs}.

Besides the action of gravitational radiation we also need to
account for dissipation. In a neutron star there are several viscosity mechanisms that could 
potentially limit or
suppress an instability. 
Here we assume a mixture of degenerate neutrons, protons and electrons where dissipation is due to the familiar shear and bulk viscosities.
This is strictly true for normal, i.e. non-superfluid matter. At temperatures $ T \lesssim 10^9\,\mbox{K}$ the likely
presence of neutron and proton superfluid condensates leads to a {\em multi-fluid} system which features
several additional shear/bulk viscosities \cite{monsterNA} and the so-called mutual friction. The latter effect is
included in our analysis but we will ignore the extra viscous degrees of freedom since they are expected to be comparable to the ones of the single-fluid system. In this case mutual friction is likely the dominant damping mechanism in the superfluid regime.

 
The nature of the standard viscosity is well known. Shear viscosity is dominated by electron (neutron) 
collisions when neutrons are superfluid (normal) \cite{cl87}. The $\beta$-equilibrium reactions responsible for bulk 
viscosity are assumed to be those of the modified Urca process \cite{STbook}. 
The associated viscosity coefficients can be found in the literature \cite{cl87,il91}; the same coefficients are used in the present work (also, a more elaborate study on shear viscosity including the dependency on the proton fraction does not change our results significantly \cite{andersson_etal2005}).

The more complex issue of the formation of a viscous boundary layer at the crust-core interface has never been discussed in the context of the $f$-mode, although it is known to be a key damping mechanism for the r-mode instability (see e.g. \cite{nakk01} and references therein). It is neglected here, firstly, because we are mostly interested in newborn neutron stars with temperatures comparable to the melting temperature of the crust ($\sim 10^{10}$ K \cite{STbook}); at this stage the crust physics is highly uncertain and, secondly, because in colder stars the instability is most likely suppressed by mutual friction (see discussion below).

Mutual friction in neutron stars is a less familiar concept. When the bulk of the interior matter is in a superfluid state, the possibility of 
large scale relative flow between the fluids is counteracted by mutual friction, which refers to the 
vortex-mediated coupling between the fluids. It can be quantified in terms of the drag parameter 
$\cR$ which is (roughly) the ratio of the drag force to the Magnus force experienced by a moving neutron vortex. 
The $f$-mode in superfluid neutron stars generically contains a `counter-moving' degree of freedom and is therefore susceptible to 
damping by mutual friction \cite{lm95,2fmode}. 


In accordance with the strategy followed for the growth timescale $t_{\rm gw}$, we calculate the shear and bulk viscosity 
damping timescales ($t_{\rm sv}$ and $t_{\rm bv}$, respectively) by using the inviscid relativistic mode eigenfunctions
in Newtonian formulae (these can be found in \cite{il91}).

Our single-fluid numerical scheme is oblivious to superfluid hydrodynamics. Yet, our analysis should somehow 
account for mutual friction since it is likely the dominant dissipative effect in `cold' neutron stars. In order to do so, 
we rely on results from calculations in Newtonian superfluid models. From the recent work in \cite{2fmode} 
we can extract the mutual friction damping timescale,
\be
t_{\rm mf} \approx \frac{ (2\ell+3)^3 (2\ell+5)}{2 (\ell+1)(3\ell+5)} 
\left ( \frac{c_s^2 R}{G M} \right )^2
\left ( \frac{1 +\cR^2}{\cR \Omega} \right ) 
\label{tmf}
\ee
where $c_s$ is the sound speed and $M,R$ are the stellar mass and radius.
This expression should be adequate for our purposes.

The combined total dissipation leads to the following expression for the
mode energy (e.g. \cite{il91,nakk01}),
\be
\dot{E}_{\rm m} = -2 E_{\rm m} \left (\frac{1}{t_{\rm gw}} + \frac{1}{t_{\rm diss}} \right )
\label{dEfull}
\ee
where the various dissipation timescales add up as ``parallel resistors'', 
$t^{-1}_{\rm diss} = t^{-1}_{\rm sv} + t^{-1}_{\rm bv} + t^{-1}_{\rm mf}$.


{\em The f-mode instability window---}
In the absence of dissipation the $f$-mode instability is active for 
any rotation $ \Omega > \Omega_\rc$ where the critical rotation is defined by $\sigma_i (\Omega_\rc) =0$. 
Relativistic stars generally enjoy lower critical frequencies (and therefore are more instability-prone)
as compared to Newtonian models \cite{sf98}. For the relativistic star considered in the following discussion, and for 
the particular $m=4$ $f$-mode multipole (which is the dominant one as we shall shortly see) the threshold for
the CFS instability is at $\Omega_\rc/\Omega_\rK \approx 0.83$.

\begin{figure}
\centerline{\includegraphics[width=0.44\textwidth]{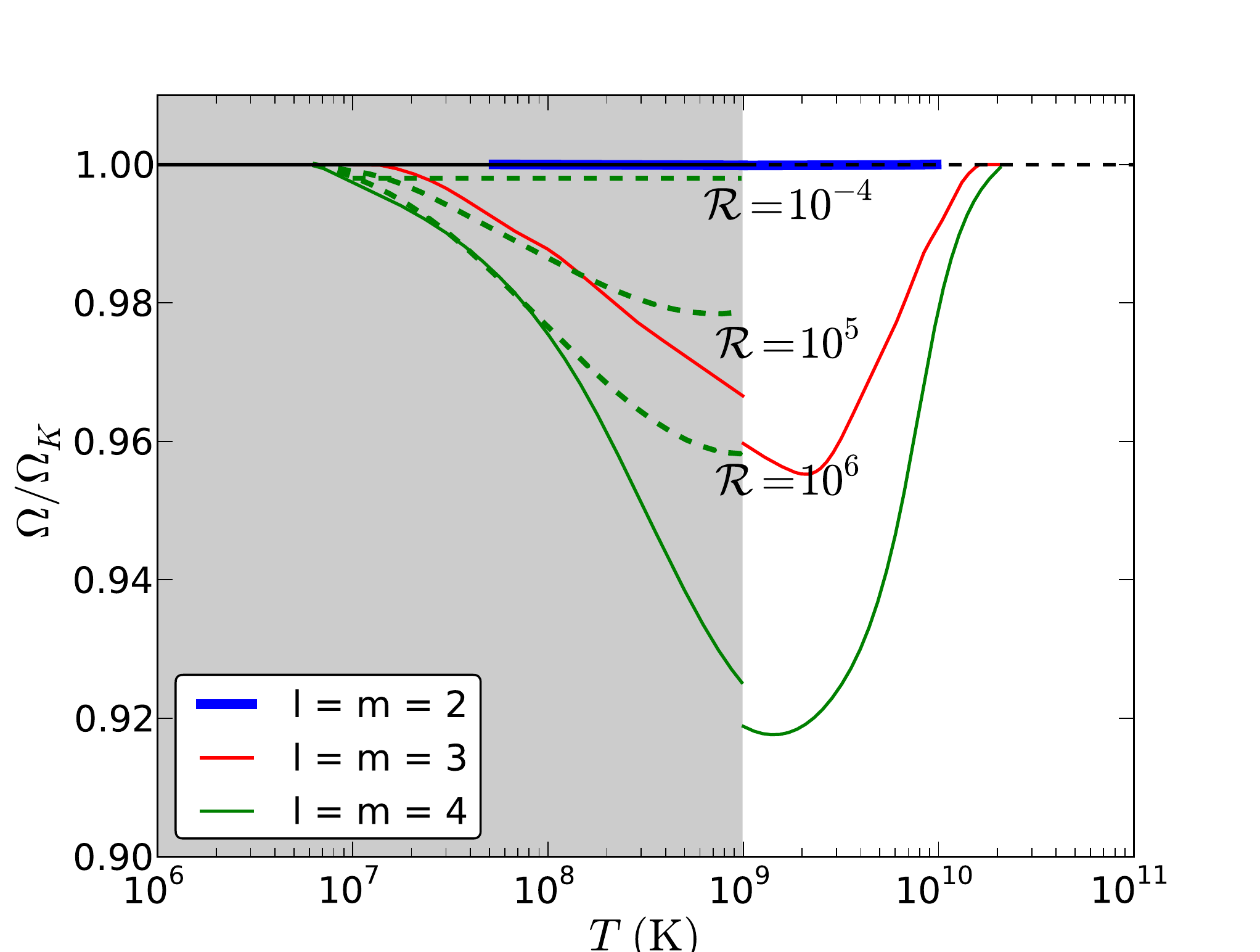}}
\caption{The relativistic f-mode instability window. The curves are the solutions
of $-t_{\rm gw} = t_{\rm diss}$, for a $N=0.73$ polytropic model with $M=1.48\,M_\odot$, $R=10.47\,\mbox{km}$ 
(in the $\Omega=0$ limit).
The shaded area indicates the presence of superfluidity (for a fiducial $T_{\rm cn} = 10^9\,\mbox{K}$). 
The dashed curves represent different choices for the mutual friction drag $\cR$ (shown only for the $m=4$
multipole).} 
\label{fig:fwindow}
\end{figure}

In the full dissipative system the boundary of the $f$-mode instability region is determined by
$\dot{E}_{\rm m} = 0$ which according to \eqref{dEfull} is equivalent to $-t_{\rm gw} = t_{\rm diss}$.
The resulting  $\Omega(T)$ instability curve is shown in Fig.~\ref{fig:fwindow} and is the main result
of this Letter. The particular example shown in the figure corresponds to a neutron star with representative values for mass and radius, approximated by
a $N= 0.73$ polytropic model which at the non-rotating limit leads to $M= 1.48\,M_\odot$ and $R= 10.47\,\mbox{km}$.
For this model the Kepler frequency is $ f_\rK = \Omega_\rK /2\pi = 1.093$ kHz.

According to Fig.~\ref{fig:fwindow}, the $m=4$ multipole is the one associated with the 
widest instability window. This mode's inertial frame frequency is $f_i = \sigma_i/2\pi = 640-2290\,\mbox{Hz}$, corresponding
to the instability region $\Omega/\Omega_\rK \approx 0.92 - 1$. The next lower multipole $m=3$ is unstable for $\Omega/\Omega_\rK \approx 0.96 - 1$,
spanning the frequency range $f_i = 200-1140\,\mbox{Hz}$. In contrast, the instability window for the $m=2$ mode is negligibly small, as it
is the case in Newtonian calculations \cite{il91}. It should be pointed out, however, that a finite-size $m=2$ instability window is present in 
models with stiffer, but not unrealistic, equations of state.

As anticipated, the $f$-mode instability is blocked by bulk viscosity in the high temperature
regime $T \gtrsim 2\times 10^{10}\,\mbox{K}$. At lower temperatures the instability has to 
compete against shear viscosity and vortex mutual friction. The former effect suppresses the instability below 
$\sim 10^7\, \mbox{K}$. The impact of mutual friction is rather more severe. The instability is completely suppressed 
below a critical temperature $T_{\rm cn} \sim 10^9\,\mbox{K}$ for neutron superfluidity in the core. 
This result assumes `standard' mutual friction, i.e. scattering of 
electrons by the magnetised neutron vortices, in which case $\cR \approx 10^{-4}$ \cite{als}. 
It is possible, however, that mutual 
friction is dominated by the interaction between
vortices and 
quantised fluxtubes.
This interaction could lead to vortex pinning; formally, this would correspond to
$\cR \gg 1$, where it is unclear what the true strength of $\cR$ is.
It is likely that the assumed mutual friction force 
leading to \eqref{tmf} is not applicable in the case of direct vortex-fluxtube interactions. If we
treat $\cR$ as a free parameter, we find that mutual friction has a negligible effect on the instability
for $\cR \gtrsim 10^{8}$.

\begin{figure}
\centerline{\includegraphics[width=0.44\textwidth]{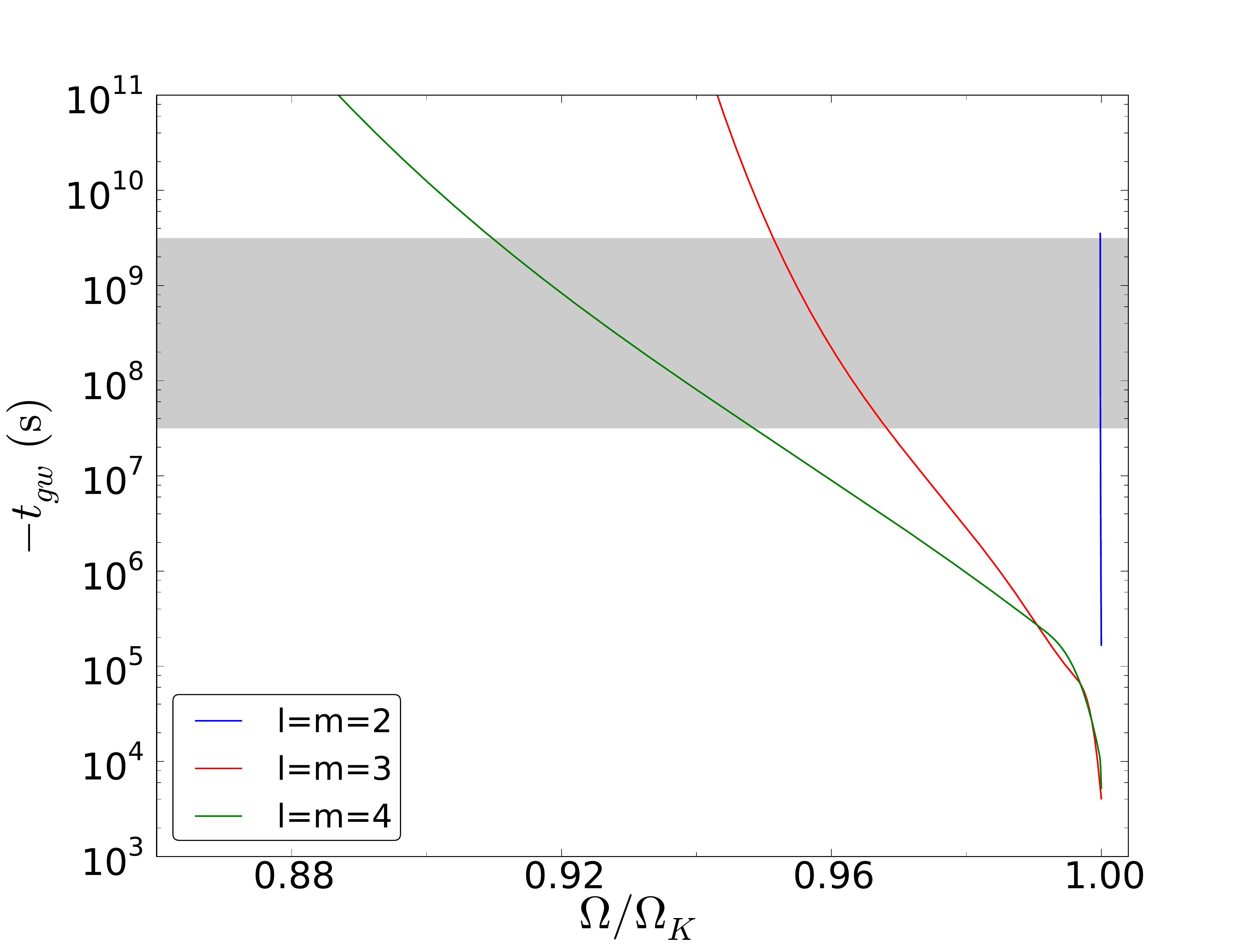}}
\caption{The $f$-mode growth timescale for the lowest three multipoles as a function of the stellar frequency. 
The stellar model is the same as in Fig.~\ref{fig:fwindow}. The shaded area represents the range of the cooling timescale 
\eqref{coolaw}, from an initial temperature  $T = 5\times 10^{10}\,\mbox{K}$ to a final 
$T = T_{\rm cn} = (5-9) \times 10^8 \,\mbox{K}$.}
\label{fig:tgw}
\end{figure}

Not surprisingly, the overall shape of the relativistic instability window does not differ much from the one of Newtonian models \cite{il91,lm95}, although it is clearly larger.
What is 
different is the instability's growth timescale $t_{\rm gw}$. 
Numerical results for the growth rate of the first few multipoles are displayed in Fig.~\ref{fig:tgw}. 
Focussing on the $m=4$ and $m=3$ modes we find a timescale $t_{\rm gw} \approx 10^3-10^5\,\mbox{s}$ at 
$\Omega \approx \Omega_\rK$. Both modes grow (almost) equally fast above $0.98\,\Omega_\rK$, however the $m=3$ mode may be 
more detectable due to its lower $f_i$. The growth timescale is a factor $10 - 100$ {\em shorter} than the corresponding Newtonian 
results \cite{il91, andersson2003}. This result, which can be attributed to the lower relativistic $\Omega_c$, is clearly good news for the gravitational wave observability of the $f$-mode.  

The astrophysical systems in which the $f$-mode instability is more likely to operate are 
newborn neutron stars, spinning with $\Omega \gtrsim 0.9\,\Omega_K$ and with a temperature 
$T \lesssim 2 \times 10^{10}\,\mbox{K}$. Once the star is inside the instability region, 
the efficiency of the $f$-mode as a source of gravitational radiation is largely decided by the 
ratio $t_\rc/t_\rgw$ where $t_\rc$ is the cooling timescale. The standard cooling law due to the 
modified Urca process is \cite{STbook},
\be
t_\rc (T) \approx \left [\, (T/10^9\,\mbox{K} )^{-6} - ( T_i/10^9\,\mbox{K})^{-6} \, \right )\, \mbox{yr}
\label{coolaw}
\ee
where $T_i$ is the initial temperature. Setting $T_i = 2 \times 10^{10}\,\mbox{K}$ and 
$T=T_{\rm cn} \approx (5-9)\times 10^8 \,\mbox{K}$ (as suggested by the cooling of the
neutron star in Cassiopeia A \cite{CasA}) we obtain $t_\rc \approx 1 - 100\,\mbox{yr}$. 
Thus, the most favourable scenario where $t_{\rm gw} \ll t_\rc$ is clearly relevant, in which case the mode has ample time 
to grow and eventually saturate by non-linear physics, while emitting gravitational waves.  

It is conceivable that newborn neutron stars spinning near $\Omega_{\rm K}$ are also endowed with strong magnetic fields. In this case, the electromagnetic spin-down could operate on a timescale comparable to the $f$-mode growth. This leads to a timescale roughly given by $\Omega/\dot{\Omega} \sim 10^{3}\,(\Omega_K/\Omega)^{2} ((B/10^{15}\,\mbox{G}))^{-2}$ s. Thus, a magnetar-strong magnetic field could effectively choke the instability.


{\em Concluding discussion---} The results presented in this Letter 
add more strength to the possibility that CFS-unstable $f$-modes in fast spinning newborn neutron stars 
could be a promising source of gravitational waves.
We should, of course, emphasize the obvious caveat that the required near-Kepler birth spin rates are only expected 
in a small fraction of newborn neutron stars. Our results should serve as the stepping stone towards a revisit of the 
gravitational wave observability of realistic, rapidly spinning neutron stars. We conclude with a discussion of 
some key issues that should be addressed by future work.

Our analysis made use of a polytropic stellar model as an approximation to
realistic neutron stars. Future calculations should explore a wide range of realistic 
equations of state. Although we do not anticipate any qualitative change in our conclusions, 
some moderate variation in the mode's frequency $f_i$ 
and growth timescale $t_{\rm gw}$ is to be expected among different equations of state. 

Another improvement on our work would be to abandon the Cowling approximation and study
oscillations of rapidly rotating neutron stars in full General Relativity. Recent work
along this line suggests that the critical rotation $\Omega_\rc$ for the onset of the CFS instability is 
pushed to lower values \cite{zink10}, a clear indication of a stronger $f$-mode instability, particularly for
the lower-order multipoles. 

The instability could be also enhanced if the stellar rotation departs from that of a rigid body. 
Such differentially rotating stars are known to have a larger instability parameter space as compared 
to the uniformly rotating models considered here \cite{lai95,kruger10}. A key related issue is
to determine whether differential rotation survives long enough as to allow for a significant mode growth.    
 
It would be also desirable to gain a better understanding of $f$-mode damping mechanisms. 
At the level of linear perturbations, it would be worth exploring how the instability window is modified if exotic matter 
(deconfined quarks, hyperons) is present in the neutron star core. Although the recent discovery of a $2\,M_\odot$-neutron star \cite{demorest}
leaves little room for a hyperonic composition, the presence of strange quark matter in the color-flavor-locked
state remains an open (and exciting!) possibility \cite{alford05}.

At the non-linear level, it is crucial to set an upper limit on the maximum amplitude of the instability. This would have
obvious implications for the gravitational wave observability because the mode could saturate as a result of
non-linear couplings with other modes. 
In this respect, a recent non-linear study is quite optimistic, 
suggesting observable signals from as far as 10 Mpc for Advanced LIGO and VIRGO \cite{kastaun10}.
Another different non-linear effect that could cause mode saturation
is non-linear bulk viscosity in the so-called suprathermal limit \cite{alford10}.

Finally, future modeling should account for the presence of a crust
and its impact on the $f$-mode instability. This could include the
damping due to a viscous boundary layer at the crust-core interface,
and even the direct breaking of the crust by a large amplitude $f$-mode
with the subsequent dissipation of energy at the fracture sites \cite{friedman83, ushomirsky2000}.
We hope to address some of these exciting issues in the near future.



This work was supported by the German Science Foundation (DFG) via SFB/TR7. EG was supported by a VESF fellowship, KG acknowledges support from  an Alexander von Humboldt fellowship. We thank the referees for helpful suggestions to improve this article. We also thank A.~Passamonti for pointing out an error in the original version of this paper.

\bibliographystyle{apsrev4-1}
\bibliography{references}

\end{document}